\documentclass[11pt,preprint2]{aastex}

\shorttitle{Limits on Planets Around White Dwarf Stars}
\shortauthors{Mullally et al.}

\begin{document}


\title{Limits on Planets Around Pulsating White Dwarf Stars}


\author{Fergal Mullally\altaffilmark{1}, D. E. Winget\altaffilmark{2}, Steven Degennaro\altaffilmark{2}, Elizabeth Jeffery\altaffilmark{2}, S.~E.~Thompson\altaffilmark{3}, Dean Chandler\altaffilmark{4}}
\and \author{S. O. Kepler\altaffilmark{5}}


\altaffiltext{1}{Department of Astrophysical Sciences, Princeton University, Princeton, NJ 08544}
\altaffiltext{2}{Department of Astronomy, University of Texas at Austin, Austin TX 78712}
\altaffiltext{3}{Department of Physics and Astronomy, University of Delaware, Newark, DE 19716}
\altaffiltext{4}{Meyer Observatory, Clifton, TX 76530}
\altaffiltext{5}{Instituto de F\'{i}sica, Universidade Federal do Rio Grande do Sul, 91501-900 Porto-Alegre, RS, Brazil}



\begin{abstract}

We present limits on planetary companions to pulsating white dwarf stars. A subset of these stars exhibit extreme stability in the period and phase of some of their pulsation modes; a planet can be detected around such a star by searching for periodic variations in the arrival time of these pulsations. We present limits on companions greater than a few Jupiter masses around a sample of 15 white dwarf stars as part of an on-going survey. One star shows a variation in arrival time consistent with a 2\,$\mathrm{M_J}$ planet in a 4.5 year orbit. We discuss other possible explanations for the observed signal and conclude that a planet is the most plausible explanation based on the data available.

\end{abstract}



\section{Introduction \label{intro}}

%

All main-sequence stars with mass less than about 8\,M$_{\odot}$ will end their lives as white dwarf stars (WDs). As such, WDs are a fossil record of star formation in the Galaxy from earliest times to just a few hundred million years ago. WDs offer a window into the ultimate fate of planetary systems, including our own solar system, and whether planets can survive the final stages of stellar evolution.
The properties of a WD are relatively insensitive to the mass of the progenitor: the mass distribution of isolated WDs is narrowly distributed in a peak 0.1\,M$_{\odot}$ wide around a mean mass of 0.59\,M$_{\odot}$ \citep{Kepler07}. Surveys of WDs can therefore search a wide range of main sequence progenitor masses and the low luminosity of a WD means any companion planets can be potentially followed up with current direct detection technology in the mid-infrared.

%
\citet{Livio84} considered the fate of a planet engulfed in the envelope of a red giant star. They determined that below a certain mass the planet would be evaporated and destroyed, but larger objects would accrete material and spiral in toward the stellar core. They predicted that the end state of these systems would be a tight binary consisting of a white dwarf and a brown dwarf and suggested this mechanism might explain the origin of cataclysmic variable (CV) systems.

Planets that are not engulfed, and sufficiently far from the stellar surface that tidal drag is small, will drift outwards to conserve angular momentum \citep[as described in][]{Jeans24}. \citet{Duncan98} investigated the stability of the outer solar system when the sun undergoes mass loss as a red giant. They found that the system was stable on timescales of at least 10 billion years for reasonable amounts of mass loss, but for larger amounts typical of WDs formed from more massive stars, the planets' orbits became unstable on timescales of $\lesssim 10^{8}$ years. \citet{Debes02} looked what would happen if the orbits of two planets became unstable and determined that if orbits crossed, the most likely result was that one planet would be scattered into a shorter period orbit, while the other would be boosted into a longer orbit or ejected from the system. For planets scattered inward, the extreme flux from a newborn white dwarf would strip the outer atmospheric layers. \citet{Villaver07} estimated that a 2\,$\mathrm{M_J}$ planet 1.8\,AU from a young white dwarf would lose half its mass in this manner. The picture drawn by this brief survey of theory is a population of planets in long period orbits around WDs, with a number of objects scattered closer to the star, or in very tight binaries.
	
Searches for sub-stellar companions to WDs have concentrated on exploiting the lower contrast between star and companion, especially in the infrared \citep[e.g.][]{Probst83, Burleigh02, Farihi05survey, Friedrich06, Debes06, Mullally07}. Although a couple of brown dwarf stars have been found with this approach \citep{Zuckerman92, Farihi04}, as yet no direct detections of planets have been claimed. \citet{Silvotti07}, using the same timing method discussed here, report the detection of a planetary mass companion in a 1.7\,AU orbit around an extreme horizontal branch sdB star.
 
Pulsating WDs allow the possibility of searching for the presence of companion planets as changes in the observed arrival time of the  pulsations. Hydrogen atmosphere (DA) WDs pulsate in an instability strip approximately 1200\,K wide near 12,000\,K and are known as DAVs (or ZZ~Ceti stars). The pulsations are non-radial g-modes with periods of order 100--1500s and amplitudes of a few percent.

The pulsation properties of DAVs vary with temperature. Those near the hot end of the strip tend to have a smaller number of shorter period, lower amplitude modes with sinusoidal lightcurves and are known as hDAVs, while those nearer the red edge show more, larger amplitude modes and asymmetric pulse shapes. The change in pulsation properties is most likely due to increasing depth of the convection layer near the surface of the star \citep[][and subsequent articles]{Brickhill83}.

\citet{Stover80} first noted that modes on hDAVs often exhibited an impressive stability in the period and phase of pulsation. \citet{Kepler05a} measured the rate of period change, $\dot{P}$, of one mode in the hDAV \object[WD0912+354]{G117-B15A} to be $3.57(82) \times 10^{-15}$, while \citet{Mukadam03} constrained $\dot{P}$ of a mode in \object[WD0133-116]{R548} to $\le 5.5(1.9) \times 10^{-15}$ and \citet{ODonoghue87} constrained $\dot{P}$ of \object[WD1425-811]{L19-2} to $<3.0 \times 10^{-14}$. Together with pulsars, these objects are the most stable astrophysical clocks known.


Measurements of $\dot{P}$ require datasets of between 10 and 30 years, but the investment in time yields a suitable scientific reward. Measurement of $\dot{P}$ provides a rare opportunity to directly test models of structure and composition of the core of a star \citep{Kepler05a}, constrain the current rate of change of the gravitational constant \citep{Benvenuto04}, as well as provide useful constraints on the mass of the hypothesized axion or other super-symmetric particles \citep{Isern92, Corsico01, Bischoff-Kim07b}

If a planet is in orbit around a star, the star's distance from the Sun will change periodically as it orbits the center of mass of the planetary system. If the star is a stable pulsator like a hDAV, this will cause a periodic change in the observed arrival time of the otherwise stable pulsations compared to that expected based on the assumption of a constant period. The change in arrival time, $\tau$, is given by 
\begin{equation}
 \tau = \frac{a_p m_p \sin{i}}{M_* c}
\end{equation}

\noindent
where $a_p$ is the semi-major orbital axis of the planet, $m_p$ is the planet mass, $M_*$ is the mass of the white dwarf, $c$ is the speed of light and $i$ is the inclination of the orbit to the line of sight. In common with astrometric methods, the sensitivity increases with the orbital separation, making long period planets easier to detect given data sets with sufficiently long baselines. 

In 2003 we commenced a pilot survey of a small number of DAVs in the hope of detecting the signal of a companion planet. We present here a progress report of the first 3--4 years of observations on 12 objects, as well as presenting limits around 3 more objects based partly on archival data stretching as far back as 1970. For one object we find a signal consistent with a planetary companion. Further observations are necessary to confirm the nature of this system. For our other objects, we can constrain the presence of planets down to a few Jupiter masses at 5\,AU, with more stringent limits for stars with archival data.

\section{Our Survey}
\citet{Kleinman04} and \citet{Eisenstein06} published a large number ($\sim$ 600) candidate DAVs with spectra taken by the Sloan Digital Sky Survey \citep{Adelman06}. A follow up survey by \citet{Mukadam04} and \citet{Mullally05} confirmed 46 of these candidates to be pulsators. We selected our targets from these two papers as well as earlier known DAVs published in the literature \citep[see][]{Bergeron95, Fontaine03}.

The ideal DAV for this survey would exhibit a number of isolated, relatively low amplitude (0.5--2\%) modes. Multiplet, or otherwise closely-spaced ($\lesssim$ 1\,s) modes are difficult to resolve in single-site data and interference between the unresolved modes makes accurately measuring the phase difficult. We selected a sample of 15 stars brighter than 19$^{\mathrm{th}}$ mag for long-term study. With the exception of R548, which has a well studied double mode, we chose stars with one isolated mode with amplitude $\gtrsim 0.5\%$. Only one star, SDSS~J221458.37$-$002511.7 has two modes suitable for study.

We have monitored this sample of stars for 4 years using the Argos Prime Focus {\sc ccd} photometer \citep{Nather04} on the 2.1m Otto Struve Telescope at McDonald Observatory. We list the observed objects in Table~\ref{omcpar}, along with the period and $\dot{P}$ of the analyzed modes. We observed each object with exposure times of 5-15s for periods of 4-8 hours per night. The exposure times are chosen to be very much shorter than the Nyquist frequency of the shortest period mode observed on the star ($\gtrsim$ 100s), and the observing time to sample many consecutive cycles of the pulsation.

We reduce our data in the manner described in \citet{Mullally05} with one improvement. Argos suffers from a fluctuating bias level but does not have an overscan region. To account for this we measure the bias from a dead column and subtract this value from both our science and dark frames before flatfielding. This is clearly not ideal, but the best approach to measuring the bias available, given that the level can vary by up to 5 {\sc dn}/pixel on timescales shorter than the exposure time. 

We perform weighted aperture photometry with a variety of apertures using the IRAF package  {\it wphot}, choosing the aperture that gives the best signal-to-noise by eye. We divide the light curve by a combination of one or more reference stars, remove points affected by cloud, fit a second order polynomial to remove the long term trend caused by differential extinction, and correct our timings for the motion of the Earth around the barycenter of the solar system using the method of \citet{Stumpff80}, accounting for all UTC leapseconds up to and including January 2006. 

We combine all the data on a star in a given month for analysis, typically 8-16 hours. We first compare the alias pattern of the peaks in the Fourier transform (FT) with a window function to identify closely spaced modes and multiplets.
It is more difficult to measure the phase of closely spaced modes ($\lesssim$ 70\,$\mu$Hz) because interference between the unresolved periodicities requires significantly more data to resolve, so we focus only on isolated modes. 

Having selected a mode for analysis, we first measure the period by fitting a sine curve using the Marquant-Levenberg non-linear least squares technique \citep{Bevington69}. We attempt to obtain between two and four accurate timings on each star per year. As we accumulate data, we re-measure the period using the entire dataset, before measuring the phase of that period in each month's data using a least squares fit. We compare the observed phase to that expected based on the assumption of a constant period and plot the result in an O-C diagram. We also check that the amplitude of a mode is stable from month to month. Varying amplitudes are a symptom of either unresolved companion periods, or an instability in the pulsation mechanism. None of the modes discussed here displayed any amplitude variability inconsistent with observational error.

\section{Results}
\subsection{GD66}
\begin{figure}
\begin{center}
    \includegraphics[angle=270, scale=.4]{f1}
    \centerline{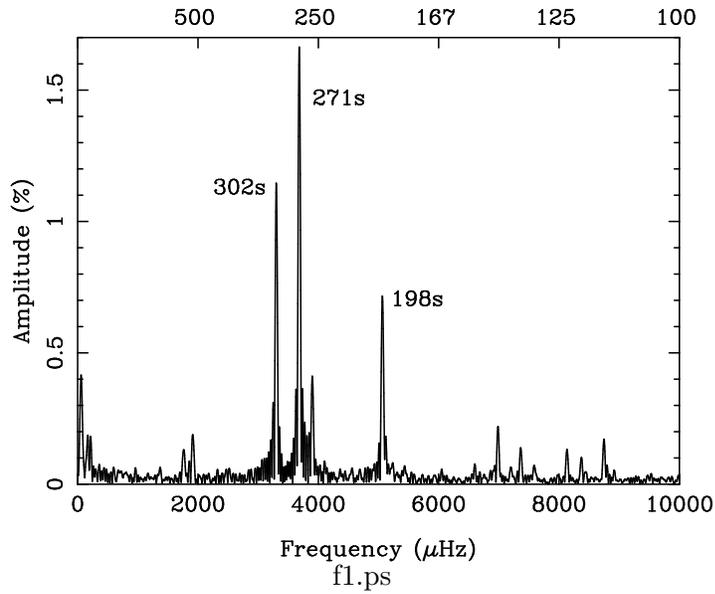}
\end{center}
\caption{Sample Fourier transform of GD66 from a single 6 hour run. The larger amplitude modes are labeled with their periods. The peaks at 271 and 198 seconds are composed of triplets of closely spaced modes separated by approximately 6.4\,$\mu$Hz that are not resolved in this data\label{gd66ft}}
\end{figure}

 \object[WD0517+307]{GD66} (WD0517+307) is an 11,980\,K, $\log{g}$=8.05, 0.64\,M$_{\odot}$ hDAV \citep{Bergeron04} with a V magnitude of 15.6 \citep{Eggen68} corresponding to a distance of about 51pc \citep{Mullally05kielb}. The FT is dominated by a single mode at 302\,s, triplets of modes separated by $\approx$6.4\,$\mu$Hz at 271 and 198 seconds and a handful of other lower amplitude modes. There are also some combination and harmonic peaks present. A sample FT is shown in Figure~\ref{gd66ft}.

\begin{figure}
\begin{center}
    \includegraphics[angle=270, scale=.4]{f2}
    \centerline{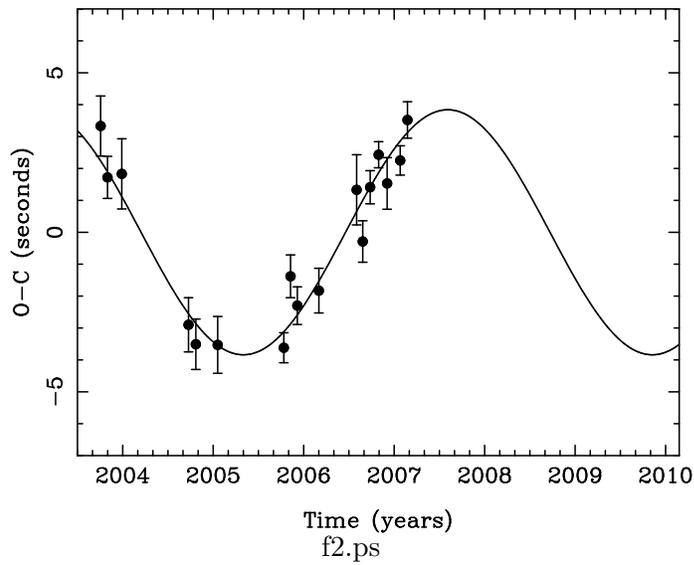}
\end{center}
\caption{O-C diagram of the 302s mode of GD66. The solid line is a sinusoidal fit to the data.  \label{gd66omc}}
\end{figure}

\begin{figure}
\begin{center}
    \includegraphics[angle=270, scale=.4]{f3}
    \centerline{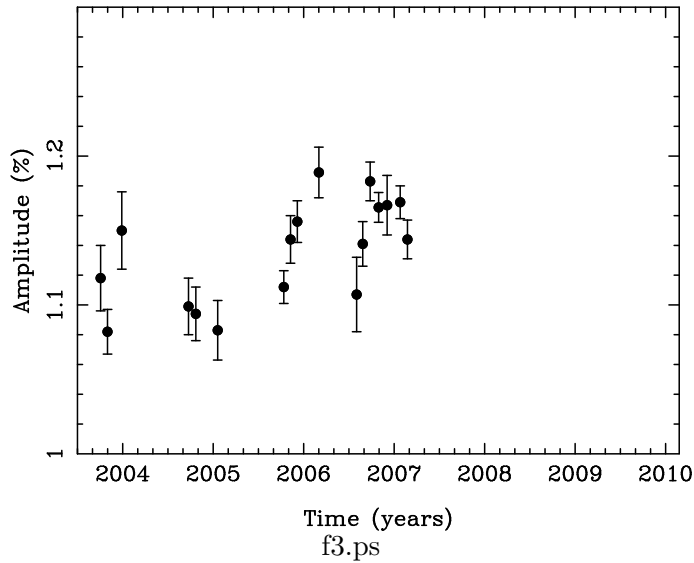}
\end{center}
\caption{Amplitude of the 302s mode of GD66 as a function of time. The error bars are the formal errors of a non-linear least squares fit, the systematic error is approximately 0.1\%. An unstable amplitude would indicate that the observed phase variations are due to some process internal to the star, however  the amplitude is stable within our ability to measure it \label{gd66amp}}
\end{figure}

The presence of closely spaced peaks at 271 and 198\,s makes it more difficult to accurately measure their phase and our analysis concentrates on the 302\,s mode. We show an O-C diagram for the arrival times of the 302\,s mode in Figure~\ref{gd66omc}. The curvature in this diagram is unmistakable. Instabilities in the pulsation modes of DAVs often manifest as variations in the amplitude of pulsation. In Figure~\ref{gd66amp} we plot the amplitude of the 302s mode as a function of time and find the amplitude varies between 1.1 and 1.2\%. We can reproduce variations of this magnitude by small changes in our reduction method, and conclude that the amplitude is stable within our ability to measure it. By fitting a sine curve to the O-C diagram we find  a period of 4.52 years and an amplitude of 3.84\,s.

The data collected to date is also consistent with a parabola. O-C diagrams of hDAVs are expected to show parabolic behavior as the cooling of the star produces a monotonic increase in the period of pulsation \citep[see][]{Kepler91}. However, based on observations of other DAVs and models of white dwarf interiors \citep{Bradley98, Benvenuto04} we expect the cooling to cause a  $\dot{P}_{\rm cool} \sim 10^{-15}$. If we fit a parabola to our data we find a $\dot{P} = 1.347(95)\times 10^{-12}$, three orders of magnitude larger than expected from cooling alone.

The tangential motion of the star with respect to the line of sight also causes a parabolic curvature in the O-C diagram. As the star moves linearly in space perpendicular to our line of sight, its distance to us changes parabolically \citep{Shklovskii70, Pajdosz95}. The USNO-B1 catalogue \citep{Monet03} quotes a proper motion of 131.6(5.0)\,mas/yr  corresponding to a $\dot{P}_{\rm pm}$ of $6.4 \times 10^{-16}$, again too small to explain the observed data.

Apparently periodic signals in O-C diagrams can be caused by random jitter or
drift in the period of the pulsator. A likelihood statistic, $\mathcal{L}$, that a given data set was caused by different combinations of observational error, period jitter and drift can be calculated according to \citet{Koen06}. We expand his methodology to determine the likelihood that the data shown in Figure~\ref{gd66omc} is the signature of a companion, or the result of stochastic changes in the pulsation period. We first calculate $\mathcal{L}$ for a model of the data that seeks to explain the data by invoking jitter or drift in the period and find values for $\log{\mathcal{L}}$ of -15.1673 and -15.0918 respectively. A model including both jitter and drift gives a similar value. 
Next, we calculate the likelihood that the residuals of the sine fit can be explained by observational error alone, and find a value of $\log{\mathcal{L}}$ of -11.5549. This strongly disfavors the hypothesis that the observations can be explained by small random changes in the pulsation period.

Time series observations of another DAV, G29-38, showed a variation in the phase of one mode over 3 months consistent with an 0.5$M_{\odot}$ object in an eccentric 109 day orbit, but a change in amplitude the following year made the mode unreliable as an accurate clock \citep{Winget90}. However, analysis of other modes on the same star failed to reproduce this behavior \citep{Kleinman94} and near-infrared imaging \citep{Kuchner98, Debes05g2938} did not detect any sub-stellar companions. It is possible that the same internal effect that mimicked a companion to G29-38 is also present in GD66 albeit with a much smaller amplitude and considerably longer period. 

If we assume the curvature is caused by a planet in a circular orbit, the best fit period is 4.52(21) years and we can use Kepler's laws to determine an orbital separation, $a_p =$ 2.356(81)\,AU. The amplitude of the sine curve, $\tau = 3.84(32)$\,s is related to the semi-major axis of the star's orbit, $a_{*}$, by $\tau c = a_{*}\sin{i}$ where $i$ is the inclination of the orbit to the line of sight and $c$ is the speed of light. The mass of the planet, $m_p$, is equal to $(M_{*}a_{*})/a_p$, where $M_{*}$ is the mass of the star. Using these two equations we find an $m_p\sin{i}$ of 2.11(14)\,M$_{J}$.  From our best fit circular orbit, we predict we will obtain observations spanning an entire orbit in early 2008.

\subsection{Other stars}
GD66 is the only star in our sample that shows strong evidence for a planetary companion. However, we can place interesting limits on the presence of planets around the other stars.

The fundamental limit on our ability to detect planets is set by the scatter in the O-C diagram. Other factors which affect this limit include the timespan and sampling pattern of the data. We perform a Monte Carlo analysis to estimate the region of the mass--orbital separation plane in which planets are actually detectable around each star based on the data available. We randomly choose a planet mass, orbital separation, eccentricity, and other orbital parameters and calculate the effect this planet would induce on the O-C diagram of a 0.59\,$M_{\odot}$ white dwarf. We then sampled this O-C diagram with the same observing pattern and error bars as our actual data for each star and fit the resulting O-C diagram with a sine curve and a parabola. If either the amplitude of the sine curve or the curvature of the parabola were measured with 3$\sigma$ confidence, the hypothetical planet was determined to be detected. We repeated this process $10^6$ times for each star and drew a shaded relief map indicating the percentage of the time a planet with a given mass and orbital separation was detected with either technique to 3$\sigma$, with dark shades indicating near 100\% detection efficiency, and white indicating regions where planets were unlikely to be detected. We show the O-C diagram and the relief map for each star in Figures 3-16. As can be seen in the figures, the annual sampling pattern means that planets with orbital periods of 1 (Earth) year are difficult to detect, resulting in the weak limits for planets at separations of slightly less than 1\,AU. 

\subsection{Notes on Individual Stars}
\begin{figure}
\begin{center}
    \includegraphics[angle=270, scale=.6]{f4}
    \centerline{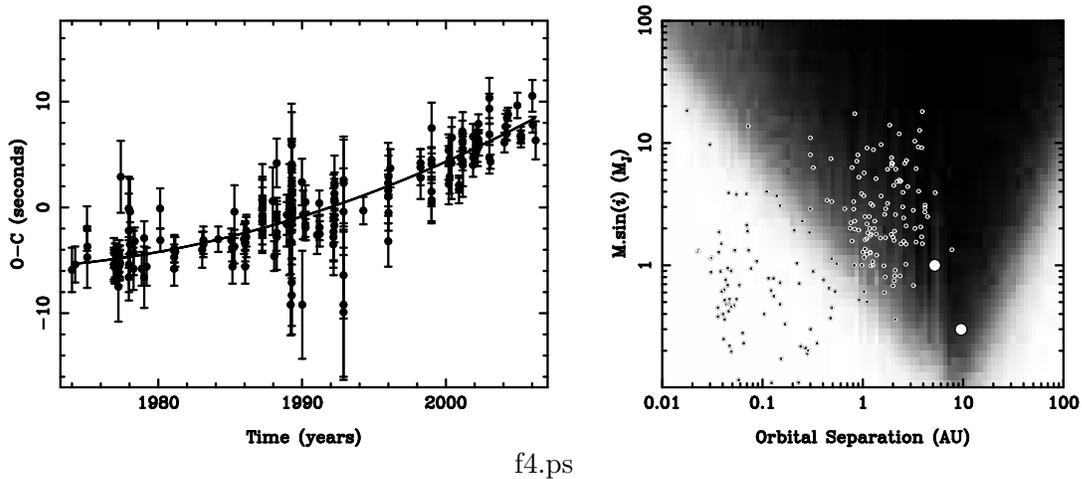}
\end{center}
\caption{{\bf Left:} O-C diagram of G117-B15A. Each point represents the phase of pulsation based on a single night of data. The solid line is a parabolic fit to the data. {\bf Right:} Relief map of the region of parameter space around G117-B15A where we are sensitive to planets. Dark regions indicate a high probability a planet would have been detected. The large white circles indicate the location of Jupiter and Saturn, while the small circles mark the postions of known extra-solar planets.\label{G117-B15A}}
\end{figure}

\noindent
\noindent
{\em G117-B15A}.--- Also known as WD0921+354. 30 years of archival data comes from \citet{Kepler05a}, although some of the more recent data in that work was taken in conjunction with this project. Where data did not come from our observations we used the O-C value quoted in their Table~1. With this timebase, the curvature caused by the change in period due to cooling becomes evident. This cooling effect is removed from the data before performing the Monte Carlo simulation. The limits on long period planets placed around this star are among the best constraints placed around any stellar object by any technique. See Figure~\ref{G117-B15A}.

\begin{figure}
\begin{center}
    \includegraphics[angle=270, scale=.6]{f5}
    \centerline{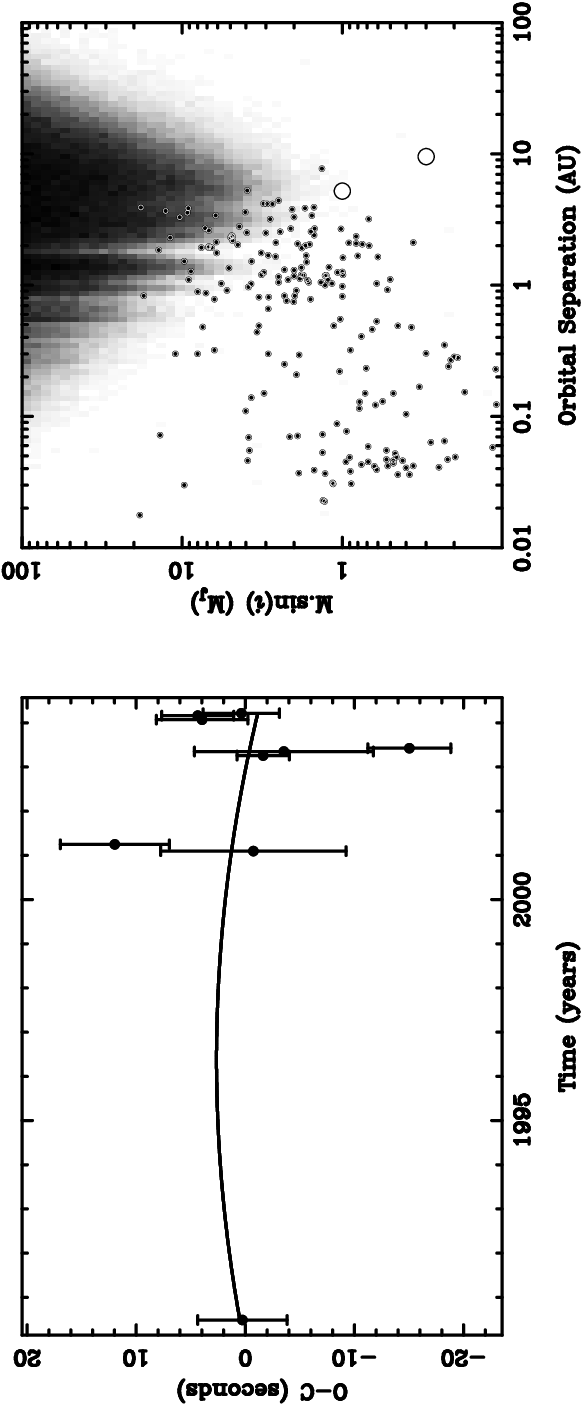}
\end{center}
\caption{Same as Figure~\ref{G117-B15A}, except for a different star, G185-32. In this O-C diagram, the points indicate the phase of pulsation measured from the combination of data taken on this star over a month or more as described in the text.\label{G185-32}}
\end{figure}

\noindent
{\em G185-32}.--- Also known as WD1935+279. The point from the early 90's comes from archival data from the Whole Earth Telescope \citep[Xcov8,][]{Castanheira04}. This extra point gives a long baseline, but the poor coverage in our survey reduces the sensitivity. See Figure~\ref{G185-32}.

\noindent
{\em R548}.--- Also known as ZZ~Ceti and WD0133-116. The entire data set comes from \citet{Mukadam03}. See Figure~\ref{R548}.

{\em SDSS J011100.63+001807.2}.--- At $g=$18.6$^{\mathrm{th}}$ magnitude, this is our faintest target and correspondingly has our weakest limits. Also, because it could only be observed under the best conditions the data coverage is quite low. The apparently impressive curvature in this O-C diagram in entirely due to the last data point
 and should be treated with considerable skepticism. See Figure~\ref{WD0111+0018}.

\noindent
{\em SDSS J135459.89+010819.3}.--- This bright (16.4$^{\mathrm{th}}$ magnitude) star has a baseline stretching back to early 2003 and some of the highest accuracy time measurements, and as a result has the best limits on planets for stars without archival data. We could have detected a Jupiter mass planet at 5\,AU had one been present. See Figure~\ref{WD1354+0108}.

\noindent
{\em SDSS J221458.37$-$002511.7}.--- This is the only star in the sample for which reasonable O-C diagrams were obtained for two modes. The O-C diagram shown (see Figure~\ref{WD2214-0025}) is the weighted sum of the O-C values for these two individual modes.


\section{Discussion}
In this pilot study we find a signal consistent with a companion planet around one star in a sample of only 15. Although further observations will be necessary to conclusively identify the cause, it augers well for the future potential of this and other white dwarf planet searches. According to the theoretical arguments discussed in \S\ref{intro}, the distribution of planets around WDs can be expected to be weighted toward planets in long period orbits. The population of hot Jupiters around the main sequence progenitors will likely be destroyed, while more distant planets will drift outward with stellar mass loss. Evidence of a planet in a relatively short orbit encourages us to continue monitoring for planets with greater orbital separations. For the other stars in our sample, we can rule out the presence of planets down to a few Jupiter masses at 5\,AU. With more data, we will extend our limits beyond 10\,AU into the regime where we expect planets to be most frequent.

Our search spans a broad range of progenitor masses. To estimate the progenitor masses we compare the spectroscopically measured \,$T_{\mathrm{eff}}$ and \,$\log{g}$ from \citet{Eisenstein06} and \citet{Bergeron04} to WD interior models from \citet{Holberg06} to obtain a WD mass. We then used the initial final mass relations (IFMRs) of \citet{Williams04} and \citet{Ferrario05} to calculate the progenitor mass, taking the weighted mean of the two relations as the best value, and adding the difference between the two methods in quadrature to the uncertainty. We present the masses in Table~\ref{ifmr}. Although there is still considerable uncertainty in this relation, the results give some indication of the type of star that created the WD.  The progenitor masses correspond to a range of spectral type from approximately B6 to F9 \citep{Habets81}, a range that is largely complementary to the radial velocity method.

\section{Conclusion}
We present our results on an on-going survey for planets around 15 pulsating white dwarf stars. Our survey data spans 3--4 years with archival data on some stars stretching back to 1970. We are already sensitive to planets down to a few Jupiter masses at distances of approximately 5\,AU. For one star, we observe a curvature in the O-C diagram consistent with a planet in a 4.5 year orbit. Further observations are necessary to span a full orbit of this candidate object. If confirmed, this will be the first planet discovered around a WD, and together with the planet discovered around an sdB, suggests that planets regularly survive the death of their parent star, and that WDs will be fruitful targets for planet searches.

\acknowledgments
This work was supported by a grant from the NASA Origins program, NAG5-13094 and is performed in part under contract with the Jet Propulsion Laboratory (JPL) funded by NASA through the Michelson Fellowship Program. JPL is managed for NASA by the California Institute of Technology.



\onecolumn

\clearpage

%
%
\begin{deluxetable}{llllr}
\tablewidth{0pt}
\tabletypesize{\footnotesize}
\tablecaption{Modes used to construct O-C diagrams \label{omcpar}}

\tablehead{
    \colhead{Star}&
    \colhead{Period}&
    \colhead{Amplitude}&
    \colhead{T$_0$}&
    \colhead{$\dot{P}$}\\
    \colhead{ }&
    \colhead{(sec)}&
    \colhead{\%}&
    \colhead{(bjd)}&
    \colhead{ }

}

\startdata
\object{G117$-$B15A}\ldots & 215.1973888(12) & 1.9 & 2442397.9194943(28) & $-1.07(49)\times10^{-13}$\\
\object{G185$-$32} & 370.2203552(55) & 0.1 & 2453589.6557652(39) & $-0.5(1.0)\times10^{-13}$\\
\object{G238$-$53} & 122.1733598(38) & 0.2 & 2453168.6334567(35) & $-5.7(2.4)\times10^{-13}$\\
\object{GD244} & 202.9735113(40) & 0.4 & 2452884.8712580(31) & $0.2(2.8)\times10^{-13}$\\
\object{GD66} & 302.7652959(21) & 1.2 & 2452938.8846146(28) & $1.347(95)\times10^{-12}$\\
\object{R548} & 212.76842927(51) & 0.4 & 2446679.833986 & $1.2(4.0)\times10^{-15}$\\
\object{SDSS~J001836.11+003151.1} & 257.777859(13) & 0.6 & 2452962.6358455(41) & $9.4(9.2)\times10^{-13}$\\
\object{SDSS~J011100.63+001807.2} & 292.9445269(90) & 1.9 & 2452963.7174455(44) & $3.87(43)\times10^{-12}$\\
\object{SDSS~J021406.78$-$082318.4} & 262.277793(11) & 0.6 & 2452941.7929412(37) & $-1.5(7.5)\times10^{-13}$\\
\object{SDSS~J091312.74+403628.8} & 172.605159(15) & 0.3 & 2453024.8275265(47) & $9.6(9.8)\times10^{-13}$\\
\object{SDSS~J101548.01+030648.4} & 254.9184503(56) & 0.7 & 2453065.6152116(41) & $7.2(3.6)\times10^{-13}$\\
\object{SDSS~J135459.88+010819.3} & 198.3077098(14) & 0.6 & 2452665.9507137(33) & $-5.3(7.8)\times10^{-14}$\\
\object{SDSS~J135531.03+545404.5} & 323.9518703(69) & 2.2 & 2453082.8582407(39) & $1.39(47)\times10^{-12}$\\
\object{SDSS~J172428.42+583539.0} & 335.536871(14) & 0.6 & 2453139.8477241(37) & $1.23(85)\times10^{-12}$\\
\object{SDSS~J221458.37$-$002511.7} & 195.1406388(64) & 0.4 & 2452821.8513218(35) & $6.2(3.6)\times10^{-13}$\\
\object{SDSS~J221458.37$-$002511.7} & 255.1524057(30) & 1.3 & 2452821.8521749(35) & $1.7(2.1)\times10^{-13}$\\

\enddata
\tablecomments{T$_0$ is the time of the arbitarily defined zeroth pulse and is given in units of barycentric corrected julian day. Data on R548 comes from \citet{Mukadam03} who do not provide a value for uncertainty in T$_0$. Except for GD66, we do not claim statistical significance for the measurement of $\dot{P}$ for any star.}
\end{deluxetable}



\begin{deluxetable}{lcrrrr}
\tablewidth{0pt}
\tablecaption{Stellar Parameters \label{ifmr}}

\tablehead{
    \colhead{Star}&
	\colhead{Magnitude}&
	\colhead{T$_{\mathrm{eff}}$}&
	\colhead{$\log{g}$}&
    \colhead{Initial Mass}&
    \colhead{Final Mass}\\

    \colhead{}&
    \colhead{}&
    \colhead{(K)}&
    \colhead{}&
    \colhead{($\mathrm{M_{\odot}}$)}&
    \colhead{($\mathrm{M_{\odot}}$)}\\
}

\startdata

G117$-$B15A & 15.7 & 11630 & 7.98 & 1.69(51) & 0.595(29)\\
G185$-$32 & 13.0 & 12130 & 8.05 & 2.10(44) & 0.638(32)\\
G238$-$53 & 15.5 & 11885 & 7.91 & 1.36(55) & 0.562(25)\\
GD244 & 15.6 & 11645 & 8.01 & 1.85(49) & 0.611(32)\\
GD66 & 15.6 & 11980 & 8.05 & 2.10(44) & 0.638(32)\\
R548 & 14.2 & 11894 & 7.97 & 1.65(51) & 0.591(28)\\
SDSS~J001836.11$+$003151.1 & 17.4 & 11696 & 7.93 & 1.45(54) & 0.571(25)\\
SDSS~J011100.63$+$001807.2 & 18.8 & 11507 & 8.26 & 3.33(37) & 0.769(37)\\
SDSS~J021406.78$-$082318.4 & 17.9 & 11565 & 7.92 & 1.40(55) & 0.566(26)\\
SDSS~J091312.74$+$403628.8 & 17.6 & 11677 & 7.87 & 1.17(60) & 0.542(25)\\
SDSS~J101548.01$+$030648.4 & 15.7 & 11584 & 8.14 & 2.62(37) & 0.693(32)\\
SDSS~J135459.89$+$010819.3 & 16.4 & 11658 & 8.01 & 1.85(49) & 0.611(32)\\
SDSS~J135531.03$+$545404.5 & 18.6 & 11576 & 7.95 & 1.61(61) & 0.580(49)\\
SDSS~J172428.42$+$583539.0 & 17.5 & 11544 & 7.90 & 1.31(57) & 0.556(25)\\
SDSS~J221458.37$-$002511.7 & 17.9 & 11439 & 8.33 & 3.75(37) & 0.814(32)\\
\enddata
\tablecomments{Sloan magnitudes are in the $g$ filter, V magnitudes for the
other stars are taken from Simbad. The sources of the temperatures, gravities
and masses are discussed in the text.}
\end{deluxetable}

\clearpage

%
%

\begin{figure}
\begin{center}
    \includegraphics[angle=270, scale=.6]{f6}
    \centerline{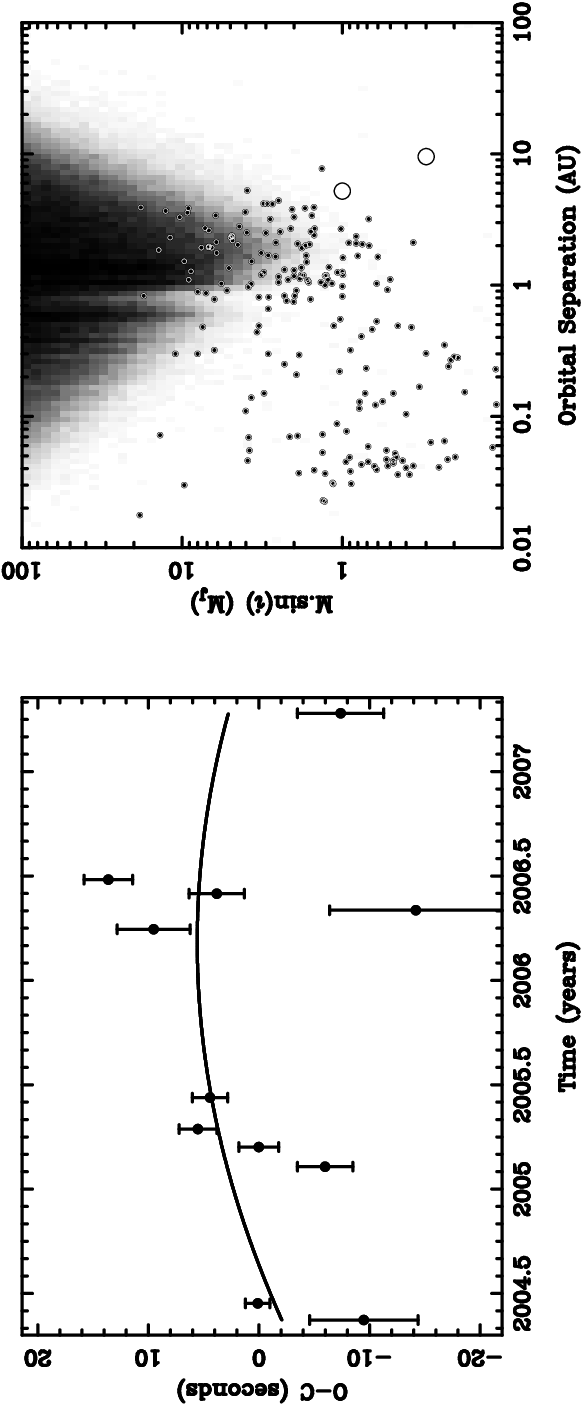}
\end{center}
\caption{G238-53\label{G238-53}}
\end{figure}

\begin{figure}
\begin{center}
    \includegraphics[angle=270, scale=.6]{f7}
    \centerline{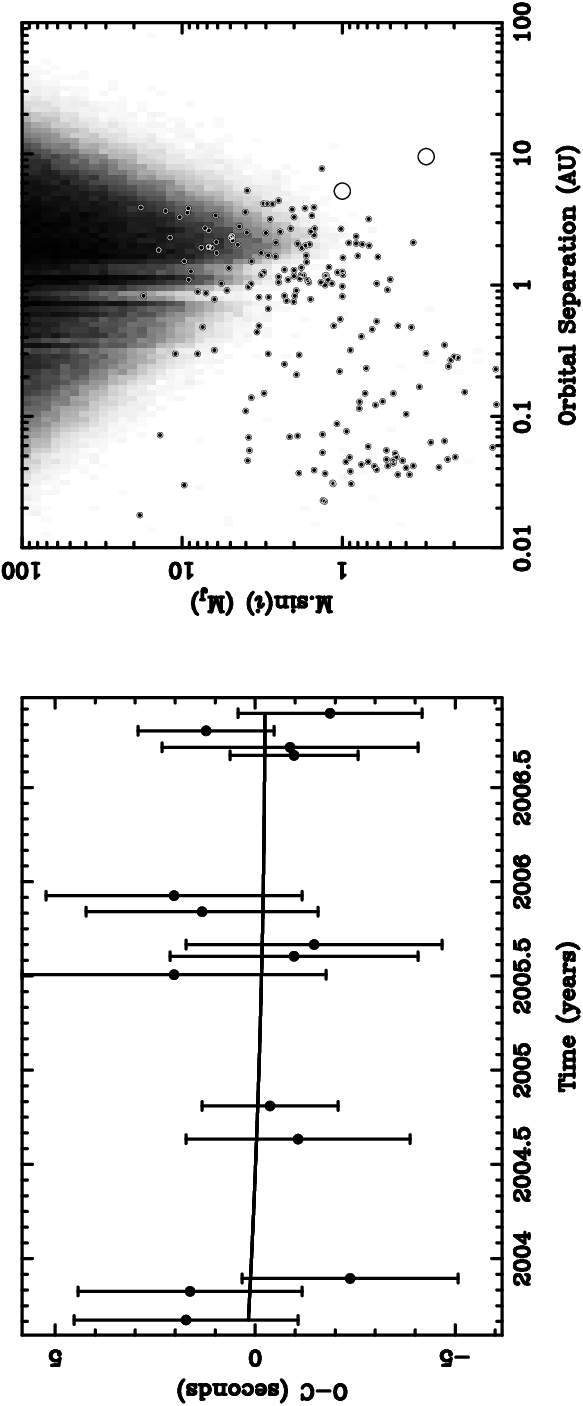}
\end{center}
\caption{GD244\label{GD244}}
\end{figure}

\begin{figure}
\begin{center}
    \includegraphics[angle=270, scale=.6]{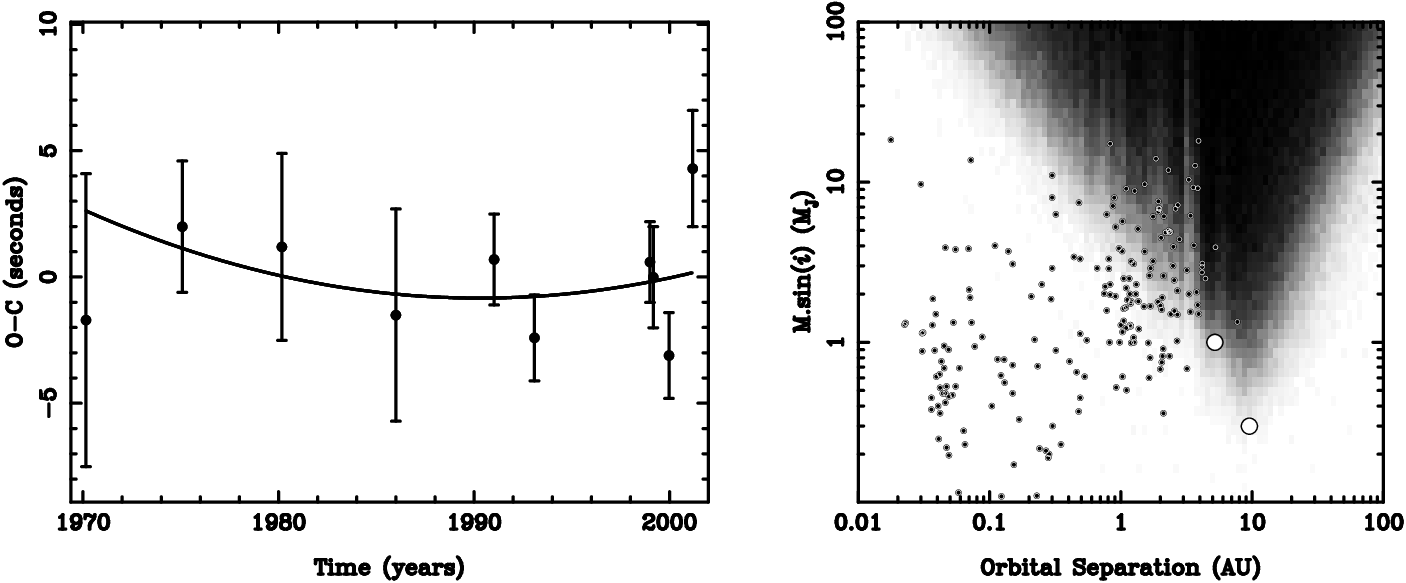}
    \centerline{f8}
\end{center}
\caption{R548\label{R548}}
\end{figure}

\begin{figure}
\begin{center}
    \includegraphics[angle=270, scale=.6]{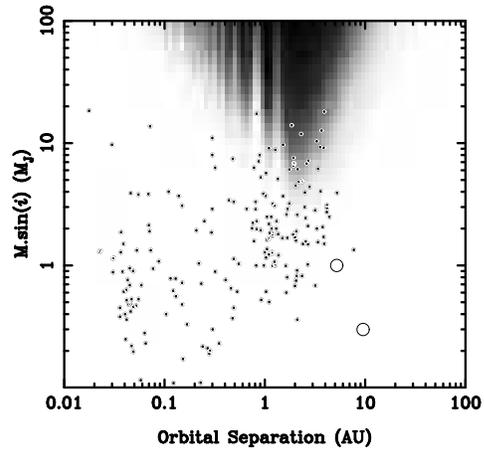}
    \centerline{f9}
\end{center}
\caption{SDSS~J001836.11+003151.1\label{WD0018+0031}}
\end{figure}

\begin{figure}
\begin{center}
    \includegraphics[angle=270, scale=.6]{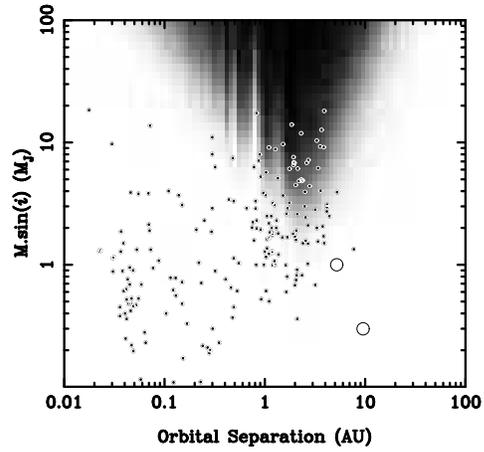}
    \centerline{f10}
\end{center}
\caption{SDSS~J011100.63+001807.2\label{WD0111+0018}}
\end{figure}

\begin{figure}
\begin{center}
    \includegraphics[angle=270, scale=.6]{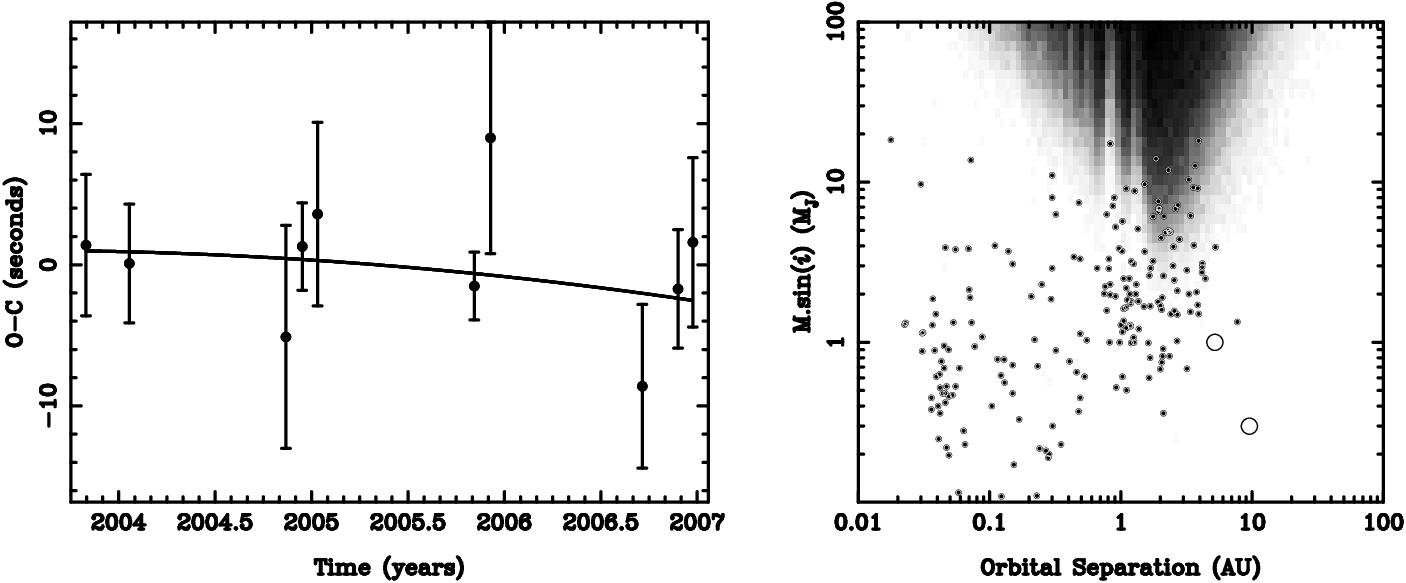}
    \centerline{f11}
\end{center}
\caption{SDSS~J021406.78$-$082318.4\label{WD0214-0823}}
\end{figure}

\begin{figure}
\begin{center}
    \includegraphics[angle=270, scale=.6]{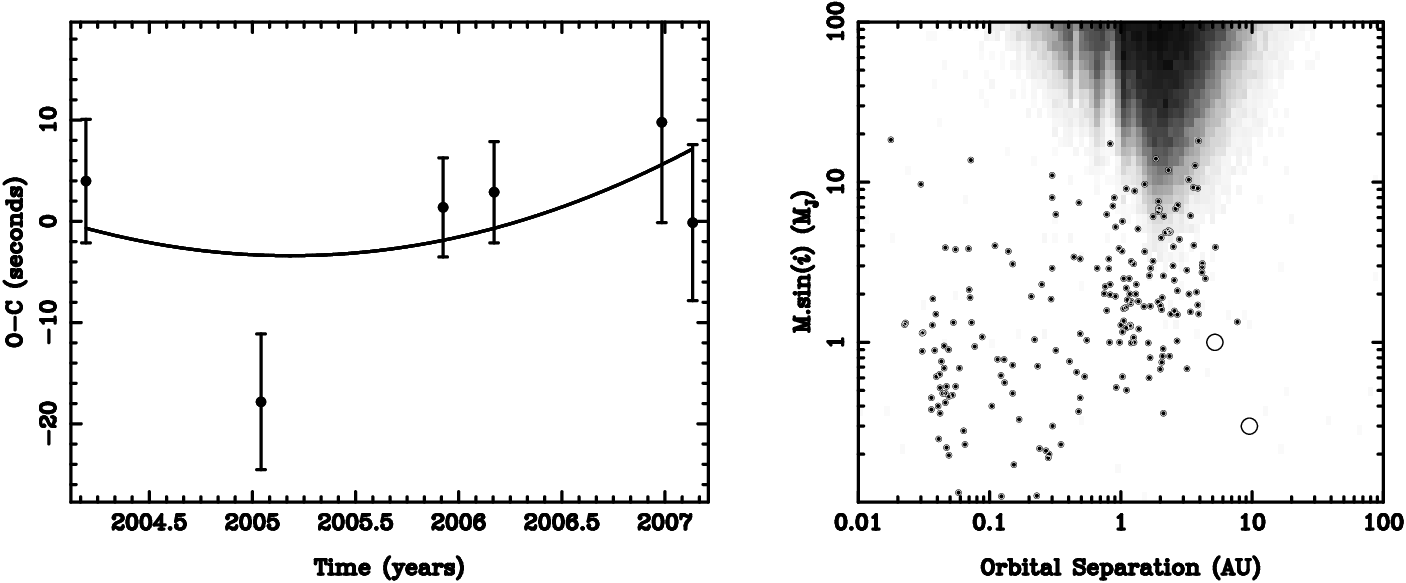}
    \centerline{f12}
\end{center}
\caption{SDSS~J091312.74+403628.8\label{WD0913+4036}}
\end{figure}

\begin{figure}
\begin{center}
    \includegraphics[angle=270, scale=.6]{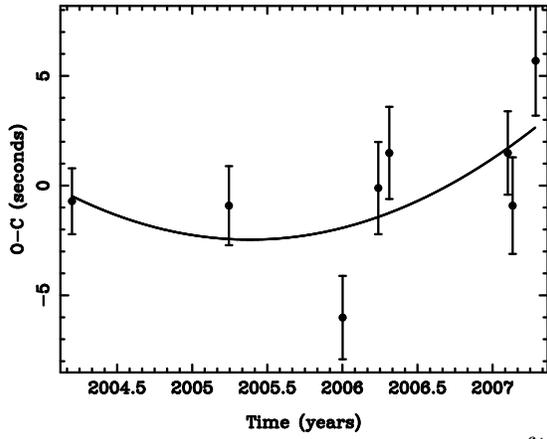}
    \centerline{f13}
\end{center}
\caption{SDSS~J101548.01+030648.4\label{WD1015+0306}}
\end{figure}

\begin{figure}
\begin{center}
    \includegraphics[angle=270, scale=.6]{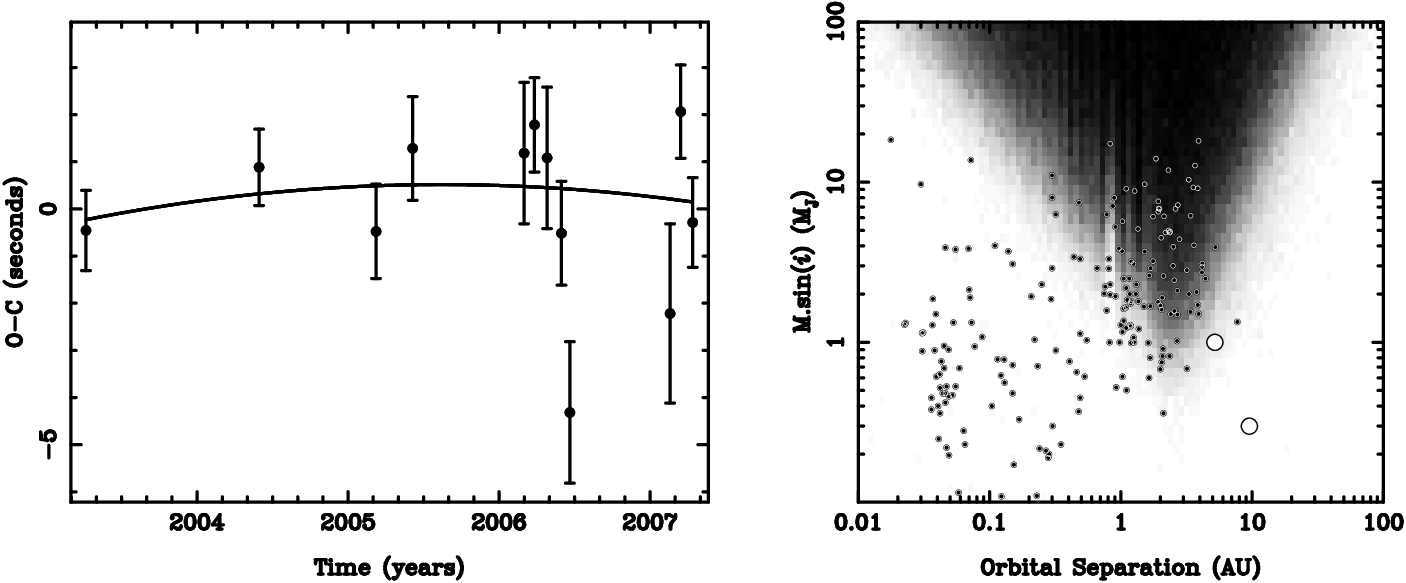}
    \centerline{f14}
\end{center}
\caption{SDSS~J135459.88+010819.3\label{WD1354+0108}}
\end{figure}

\begin{figure}
\begin{center}
    \includegraphics[angle=270, scale=.6]{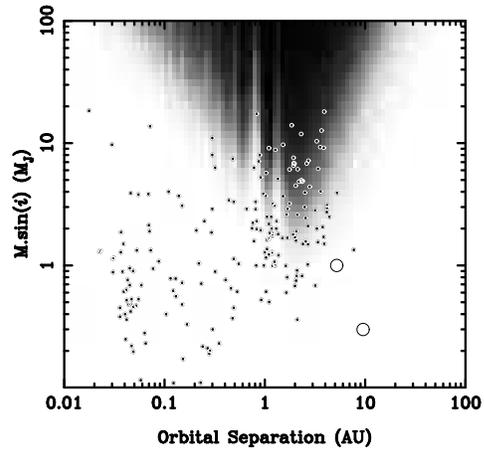}
    \centerline{f15}
\end{center}
\caption{SDSS~J135531.03+545404.5\label{WD1355+5454}}
\end{figure}

\begin{figure}
\begin{center}
    \includegraphics[angle=270, scale=.6]{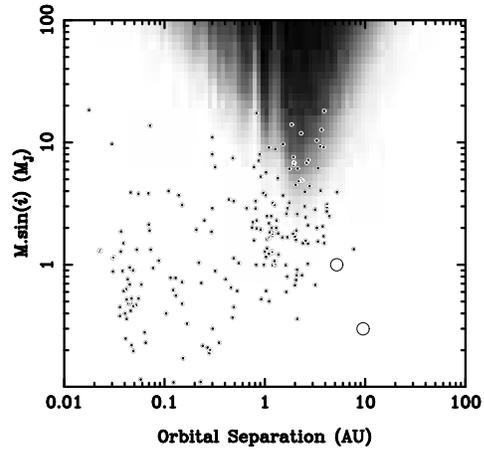}
    \centerline{f16}
\end{center}
\caption{SDSS~J172428.42+583539.0\label{WD1724+5835}}
\end{figure}

\begin{figure}
\begin{center}
    \includegraphics[angle=270, scale=.6]{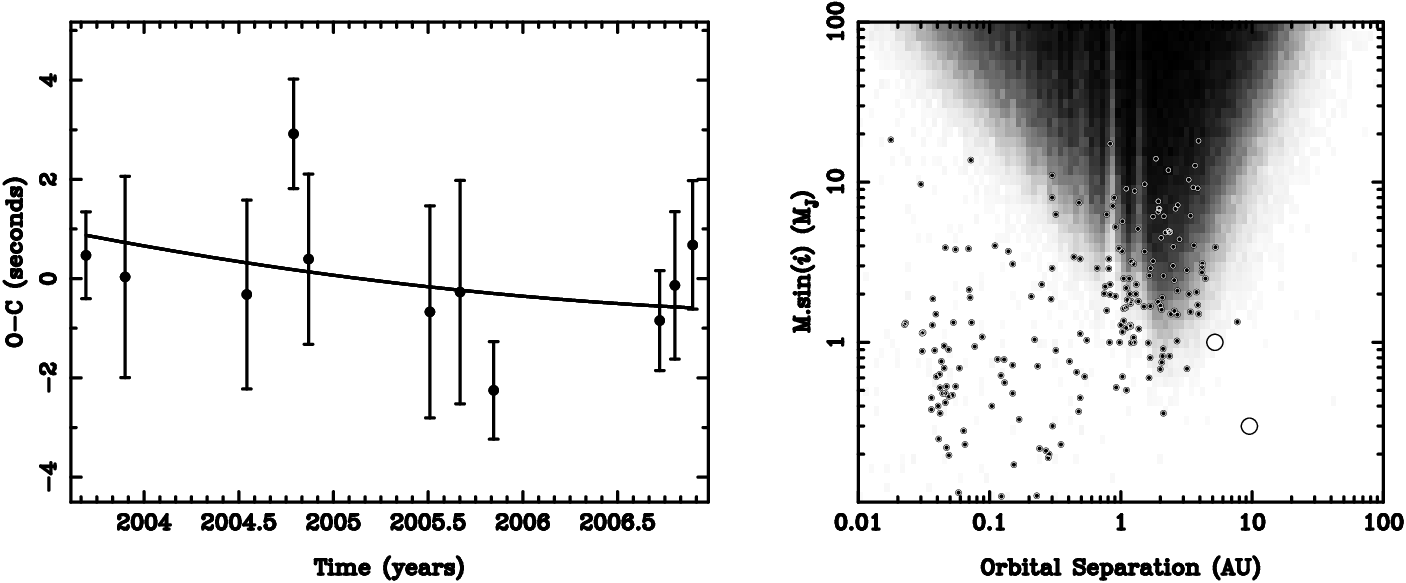}
    \centerline{f17}
\end{center}
\caption{SDSS~J221458.37$-$002511.7\label{WD2214-0025}}
\end{figure}


\end{document}